\documentclass[aps,twocolumn,showpacs,superscriptaddress,preprintnumbers,amsmath,amssymb]{revtex4}
\usepackage{graphicx}
\usepackage{dcolumn}
\usepackage{bm}
\usepackage{color}

\begin{document}

\title{Impact of intense laser pulses on the autoionization dynamics \\ of the $2s2p$ doubly-excited state of He}

\author{Anton N. Artemyev}
\affiliation{Institut f\"ur Physik und CINSaT, Universit\"at Kassel, Heinrich-Plett-Str. 40, 34132 Kassel, Germany}

\author{Lorenz S. Cederbaum}
\affiliation{Theoretische Chemie, Physikalisch-Chemisches Institut, Universit\"at Heidelberg, Im Neuenheimer Feld 229, 69120 Heidelberg, Germany}

\author{Philipp V. Demekhin}\email{demekhin@physik.uni-kassel.de}
\affiliation{Institut f\"ur Physik und CINSaT, Universit\"at Kassel, Heinrich-Plett-Str. 40, 34132 Kassel, Germany}

\begin{abstract}
The photoionization of a helium atom by short intense laser pulses is studied theoretically in the vicinity of the $2s2p\,^1P$ doubly-excited state with the intention to investigate the impact of the intensity and duration of the exciting pulse on the dynamics of the autoionization process. For that purpose, we solve numerically the corresponding time-dependent Schr\"{o}dinger equation by applying the time-dependent restricted-active-space configuration-interaction method (TD-RASCI). The present numerical results clearly demonstrate that the Fano-interferences can be controlled by a single high-frequency pulse. As long as the pulse duration is comparable to the autoionization lifetime, varying the peak intensity of the pulse enables manipulation of the underlying Fano-interference. In particular, the asymmetric profile observed for the $2s2p\,^1P$ doubly-excited state of He in the weak-field ionization can be smoothly transformed to a window-type interference profile.
\end{abstract}


\maketitle

\section{Introduction}
\label{sec:introduction}

Whenever an electronic state of a discrete spectrum is embedded in an electron continuum of the same symmetry, an ultrafast electronic decay governed by the configuration interaction effects takes place. This phenomenon is known as autoionization. Although, autoionizing resonances in atoms were observed \cite{Beutler35} and interpreted \cite{Fano35} already in 1935, a particular interest to this phenomenon arose in atomic physics with the advent of tunable synchrotron radiation sources in 1960s (see, e.g., pioneering works of Madden and Codling  \cite{MC1,MC2,MC3}). The autoionizing states manifest themselves in the photoionization (photoabsorption) cross section as prominent features with vast diversity of resonant profile shapes (see, e.g., Ref.~\cite{Sukhorukov12} for recent review on atomic autoionization).

The first theoretical description of autoionization was reported by Fano \cite{Fano61}. By considering interaction of an isolated autoionizing resonance with a single continuum, Fano has derived his well-known parametrization formula which accounts for the interference of the amplitudes for the direct and resonant (i.e., for the excitation and subsequent decay of the resonant state) photoionization pathways. According to Ref.~\cite{Fano61}, the observed shapes of the resonant profile can be described by a parameter $q$, which is determined by the ratio of the transition amplitudes for the resonant and direct ionization channels. The autoionizing resonances with $\vert q\vert \gg 1$ exhibit a peak-type profile, with  $\vert q \vert \ll 1$ a window-type shape, and with  $\vert q \vert \sim 1$ a dispersion profile. Later on, the Fano-parametrization was extended to the case of interaction of several overlapping resonances with several autoionization continua \cite{Fano65,Shore67,Shore68}.

The Fano-interference becomes particularly intricate if the autoionizing resonant state is populated by a coherent intense laser pulse which duration is comparable to the corresponding autoionization lifetime \cite{Lambropoulos80,Lambropoulos98}. Here, the exciting pulse creates a coherent superposition of the initial and resonant states which is superimposed with the autoionization continuum. If the corresponding Rabi-frequency is made larger than the autoionization width, the photoelectron spectrum exhibits a multiple-peak structure \cite{Rzaznewski81,Rzaznewski83} which is owing to dynamic interference \cite{Toyota07,Toyota08,Demekhin12a,Demekhin12b,Baghery17}. As a consequence, the profile-shape of the autoionizing resonance, observed in the weak-field limit, becomes significantly distorted. A simplified analytical model,  describing impact of the intensity and duration of the driving laser pulse on the corresponding $q$-parameter (i.e., on the shape of the Fano-profile), was formulated by Lambropoulos and Zoller \cite{Lambropoulos81}  in 1981.

The recent advent of the attosecond lasers \cite{Krausz09}, high-order harmonic generation sources \cite{Sansone06,Goulielmakis08}, and free electron lasers \cite{Ackermann07,FERMI,FLASH} awoke a renewed interest to autoionizing states. These modern high-frequency laser facilities allow one to directly address autoionizing states by a one-photon absorption. In addition, the laser pulse duration can be made comparable to the respective autoionization lifetimes. Finally, the laser pulse intensities accessed by these facilities allow to made Rabi-frequencies larger than corresponding autoionization widths. Thus, all necessary experimental conditions required for manipulation of the Fano-interference  by an exciting laser pulse, as proposed in  Ref.~\cite{Lambropoulos81}, are available.

Most of the recent experiments on autoionizing states are performed within the pump-probe scheme \cite{Fushitani08}, where a pump  high-frequency pulse prepares a transient state, and its autoionizing dynamics is then addressed and manipulated by IR or visible probe pulse. During last decade, autoionizing states of argon \cite{Wang10} and helium \cite{Loh08,Gaarde11,Holler11,Chen12,Ott13,Chini14,Blattermann14,Kaldun14,Ott14} were reinvestigated experimentally by the pump-probe scheme and theoretically by implying sophisticated models \cite{Chen13a,Chen13b} and numerical solution of the time-dependent Schr\"{o}dinger equation \cite{Argenti15}. In almost all studies mentioned above, a control over the Fano-interference has been achieved by an interplay between the intense high-frequency pump and low-frequency probe pulses.

In the present work we would like to study theoretically the possibility to manipulate Fano-interferences solely by a single  high-frequency laser pulse exciting the resonance. For this purpose, we consider autoionization of the prototypical $2s2p\,^1P$ doubly-excited state of helium atom (excitation energy is around 60~eV \cite{MC1,Silverman64,Samson94}), which persists even in He nanodroplets \cite{LaForge16}. In order to solve the time-dependent Schr\"odinger equation for a helium atom exposed to an intense coherent high-frequency laser pulse, we utilize the time-dependent restricted-active-space configuration-interaction method (TD-RASCI \cite{Hochstuhl12}). Recently, we have successfully applied this method to study dynamic interference \cite{Artemyev16} and high-order harmonic generation \cite{Artemyev17} processes in He. The essentials of the present realization of the TD-RASCI method are outlined in Sec.~\ref{sec:theory}. An impact of the pulse duration and intensity on the respective autoionization dynamics is discussed in Sec.~\ref{sec:results}. We conclude in Sec.~\ref{sec:summary} with a brief summary and outlook.

\section{Theory}
\label{sec:theory}

The total Hamiltonian governing the time-evolution of the two-electron wave function of helium atom exposed to an intense coherent linearly polarized laser pulse reads (atomic units are used throughout):
\begin{subequations}
\label{eq:HHH}
\begin{equation}
\label{eq:Htot}
\hat{H}(t)=\hat{H}_0+\hat{V}(t),
\end{equation}
\vspace{-0.5cm}
\begin{equation}
\label{eq:H0}
\hat{H}_0=-\frac{1}{2}\vec{\nabla}^2_1 -\frac{1}{2}\vec{\nabla}^2_2 -\frac{2}{r_1}-\frac{2}{r_2} +\frac{1}{\vert \vec{r}_1-\vec{r}_2\vert},
\end{equation}
\vspace{-0.25cm}
\begin{equation}
\label{eq:Hi}
\hat{V}(t)=-i\left( \nabla_{z_1}+\nabla_{z_2}\right) \mathcal{ A}_0 \,f(t) \sin(\omega t).
\end{equation}
\end{subequations}
The term $\hat{V}(t)$ describes the light-matter interaction in the dipole velocity  gauge, which is most suitable for the numerical solution of strong-field problems \cite{Cormier96,Han10}. The time-envelope of the pulse with the carrier frequency $\omega$ is determined by $f(t)$, whereas its peak intensity $I_0$ is given by the peak amplitude of the vector potential ${\cal A}_0$ (where ${\bf E} = - \partial_t{\bf A}$) by the expression $I_0 = \frac{\omega^2}{8\pi \alpha}{\cal A}_0^2$  (1 a.u. of intensity is equal to $6.43641 \times 10^{15}$~W/cm$^2$).

The spatial part $\Psi(\vec{r}_1,\vec{r}_2,t)$ of the singlet two-electron wave function of He is sought in the form of the symmetrized expansion over the two mutually-orthogonal basis sets of the one-particle functions $\{\phi_\alpha \}$ and $\{\psi_\beta \}$:
\begin{multline}
\Psi(\vec{r}_1,\vec{r}_2,t)=\sum_\alpha a_\alpha(t)\phi_\alpha(\vec{r}_1)\phi_\alpha(\vec{r}_2)\\
+\frac{1}{\sqrt {2}}\sum_{\alpha > \alpha^\prime} b_{\alpha \alpha^\prime}(t)\left[\phi_\alpha(\vec{r}_1)\phi_{\alpha^\prime}(\vec{r}_2)+ \phi_{\alpha^\prime}(\vec{r}_1)\phi_\alpha(\vec{r}_2) \right]\\
+\frac{1}{\sqrt {2}}\sum_{\alpha\beta} \left[\phi_\alpha(\vec{r}_1)\psi_{\beta}(\vec{r}_2,t)+ \psi_{\beta}(\vec{r}_1,t)\phi_\alpha(\vec{r}_2) \right].
\label{eq:tewf}
\end{multline}
The basis set  $\left\{\phi_{\alpha}(\vec{r}\,)\equiv \phi_{n\ell m}(\vec{r}\,) \right\}$ consists of the selected discrete orbitals of helium ion and describes dynamics of the electron which remains bound to the nucleus. The basis set $\left\{ \psi_{\beta} (\vec{r},t)\equiv  \psi^\alpha_{\ell m} (\vec{r},t) \right\}$ is constructed from the time-dependent wave packets and describes dynamics of the outgoing photoelectron. The employed ansatz~(\ref{eq:tewf}) restricts the present active space to configurations with only one of the electrons in a continuous spectrum and, therefore,  neglects possible double ionization of He \cite{Hochstuhl11}.

The radial parts of the one-particle basis functions $\{\phi_\alpha \}$ and $\{\psi_\beta \}$ are described in the present work by the finite-elements discrete-variable representation employing the normalized Lagrange polynomials $\left\{ \chi_{ik}(r) \right\} $ constructed over a Gauss-Lobatto grid \cite{Manolopoulos88,Rescigno00,McCurdy04,Demekhin13a,Artemyev15}:
\begin{subequations}
\label{eq:basis}
\begin{equation}
\label{eq:phibas}
\phi_{\alpha}(\vec{r}\,) =\sum_\lambda d^{\,\alpha}_\lambda\, \xi_{\lambda}(\vec{r}\,),
\end{equation}
\vspace{-0.5cm}
\begin{equation}
\label{eq:psibas}
\psi_{\beta}(\vec{r},t) =\sum_\lambda c^{\,\beta}_\lambda(t) \, \xi_{\lambda}(\vec{r}\,).
\end{equation}
\end{subequations}
with the three-dimensional basis element $\xi_{\lambda} (\vec{r}\,)$ defined as:
\begin{equation}
\label{eq:basis3d}
\xi_{\lambda} (\vec{r}\,) \equiv \xi_{ik,\ell m} (\vec{r}\,)=\frac{\chi_{ik}(r)}{r}\,Y_{\ell m}(\theta,\varphi).
\end{equation}
The explicit analytic expressions for the matrix elements of the Hamiltonian~(\ref{eq:HHH}) in terms of the normalized Lagrange polynomials $\left\{ \chi_{ik}(r) \right\} $ can be found in our previous works \cite{Artemyev16,Artemyev17}.

From Eqs.~(\ref{eq:tewf}) and (\ref{eq:psibas}) one can see that evolution of the total wave function is defined by the time-dependent vector $\vec{A}(t)$ composed of the coefficients $a_\alpha(t)$, $b_{\alpha \alpha^\prime}(t)$, and $c^{\,\beta}_\lambda(t)$. This vector was propagated according to the Hamiltonian (\ref{eq:HHH})
\begin{equation}
\label{eq:propagation}
\vec{A}(t+\delta t)=\exp\{- i P \hat{H}(t)P \delta t \}\vec{A}(t),
\end{equation}
where the one particle projector $P=1-\sum\limits_\alpha \vert \phi_\alpha \rangle \langle \phi_\alpha \vert$ acts only on the subspace of the coefficients $c^{\,\beta}_\lambda(t)$ and insures mutual-orthogonality of the two basis sets $\langle \phi_{\alpha} \vert \psi_{\beta }(t) \rangle=0$ at any time \cite{Hochstuhl11}. Equation~(\ref{eq:propagation}) was propagated by the short-iterative Lanczos method \cite{Park86}. To find the initial ground state $\vec{A}(0)$,  a propagation in the imaginary time (relaxation) with the Hamiltonian $\hat{H}_0$ (\ref{eq:H0}) has been performed starting from an arbitrary guess function.

The total autoionization width of the $2s2p\,^1P$ doubly-excited state of He is equal to about $\Gamma \approx 0.037$~eV \cite{Ho91,Scrinzi98,NgokoDjiokap11}, that corresponds to the lifetime of about $\tau \approx 18$~fs. In order to allow this state to complete its decay into the autoionization continuum $\{\psi_\beta \}$, one needs to propagate the two-electron wave function (\ref{eq:tewf}) for more than 100~fs after the end of the exciting pulse, which is time-consuming. Nevertheless, there is a more elegant way to determine the total photoionization probability and the photoelectron spectrum. It relies on the fact that the two-electron wave function $\Psi(\vec{r}_1,\vec{r}_2,t)$ contains a complete information on the final observables already right after the end of the pulse at time $T$. In order to access this information, one needs the eigenvalues $\epsilon_j$ and the corresponding two-electron eigenfunctions $\Phi_{\epsilon_j}(\vec{r}_1,\vec{r}_2)$  of the unperturbed Hamiltonian $\hat{H}_0$ (\ref{eq:H0}).

Diagonalization of the full matrix of Hamiltonian $\hat{H}_0$ (\ref{eq:H0}), constructed in the basis of the normalized Lagrange polynomials $\left\{ \chi_{ik}(r) \right\} $, is a formidable task. Nevertheless, it becomes feasible due to the use of the finite-elements discrete-variable representation of the radial coordinate \cite{Manolopoulos88,Rescigno00,McCurdy04,Demekhin13a,Artemyev15,Hochstuhl11,Hochstuhl12,Artemyev16,Artemyev17}, which results in a banded structure of the full Hamiltonian. This matrix is sparse -- the number of the non-zero elements does not exceed 0.1\%. This allowed us to employ the FEAST solver package \cite{Polizzi09,Polizzi15}, which provides accurate eigenvalues and eigenvectors of sparse matrices on a given interval of eigenvalues within reasonable computational costs. This algorithm can separately be applied to each block of the full Hamiltonian matrix with a particular symmetry  according to the quantum numbers of the total orbital angular momentum $L$ and $M$.

The total ionization yield $P_I(t)$ as a function of time is given by the projections of the two-electron wave function  $\Psi(\vec{r}_1,\vec{r}_2,t)$ onto all eigenfunctions $\Phi_{\epsilon_j}(\vec{r}_1,\vec{r}_2)$ with the energy $\epsilon_j > E_{ion}$ as
\begin{equation}
\label{eq:probability}
P_I(t)=\sum\limits_{\epsilon_j > -2}\left|\langle \Psi(t)|\Phi_{\epsilon_j} \rangle\right|^2 .
\end{equation}
Here  $E_{ion}=-2$~a.u. stays for the energy of the ground state of He$^+$. Keeping in mind that each of the projections $\left|\langle \Psi(t)|\Phi_{\epsilon_i} \rangle\right|^2$ represents a probability for population of the continuum energy interval $\Delta\epsilon_j=\frac{\epsilon_{j+1}-\epsilon_{j-1}}{2}$ centered around $\epsilon_j$, we arrive at the following expression for the photoelectron spectrum  $\sigma(\epsilon_j, t)$
\begin{equation}
\label{eq:spectrum}
\sigma(\epsilon_j, t)=\frac{\left|\langle \Psi(t)|\Phi_{\epsilon_j} \rangle\right|^2}{\Delta\epsilon_j}.
\end{equation}
Thereby, the total ionization probability (\ref{eq:probability}) and the total photoelectron spectrum (\ref{eq:spectrum}) are related as usual via  $P_I(t)=\int \sigma(\epsilon,t)d\epsilon $. Because after the end of the pulse (i.e., for $t > T$) the time-evolution of $\Psi(t)$ is governed by the Hamiltonian  $\hat{H}_0$ (\ref{eq:H0}), the observables $P_I(t>T)$ and $\sigma(\epsilon_j, t>T)$ do not evolve in time. Finally, the cubic-spline interpolation and the relation $\varepsilon_j=\epsilon_j+2$ was used to represent the photoelectron spectrum $\sigma(\varepsilon,T)$ on an arbitrary grid of the photoelectron kinetic energies $\varepsilon$.

Let us finally discuss some details of the present numerical calculations. Similarly to our previous works \cite{Artemyev16,Artemyev17}, the basis set of functions  $\{\phi_\alpha\}$, describing the dynamics of the electron which remains bound, was restricted to a set of the  hydrogen-like functions $\left\{n\ell_+\right\}$ of the He$^+$ ion. In order to arrive at a satisfactorily description of the initial ground and the intermediate doubly-excited state of He (see next section for details), all discreet wave functions with the quantum numbers $n\leq 5$ and $\ell\leq 4$ were incorporated in the $\left\{\phi_\alpha=n\ell_+\right\}$ set. The present calculations were performed for laser pulses with the sine-squared time-envelope $f(t)=\sin^2\left(\frac{\pi t}{T}\right)$. In order to investigate influence of the pulse duration $T$ on the dynamics of autoionization process, the  calculations were carried out  for different values of $T=10$, 20 and 30~fs, which are comparable to the autoionization lifetime $\tau \approx 18$~fs \cite{Ho91,Scrinzi98,NgokoDjiokap11}.

The photoelectron wave packets  $\{\psi_\beta \}$ were described on the radial grid by partial harmonics with $\ell\leq 4$. The size of the radial box $R_{max}$ was chosen to support electrons with kinetic energies of up to 100~eV during the pulse duration $T$. Our box thus supports electrons released by the one-photon ionization, as well as by the two-photon  above-threshold ionization. In particular, we select the box sizes of $R_{max}=1000$, 2000, and 3000~a.u. for the $T=10$, 20, and 30~fs pulses, respectively. The entire radial interval was divided by finite elements with the length of 2.5~a.u., each covered by 10 Gauss-Lobatto points. In order to avoid reflection of  very fast electrons from the boundary, the following mask function \cite{Bandrauk09,Krause92}
\begin{equation}
\label{eq:mask}
h(r)= \left\{\begin{array}{ll}1,& r<R_0 \\ \left(\cos \left[ \frac{\pi}{2}  \frac{r-R_0}{R_{max}-R_0} \right]\right)^\frac{1}{8},& R_0<r<R_{max}, \end{array} \right.
\end{equation}
with $R_0=R_{max}-100$~a.u., was applied to the photoelectron wave packets. Care was taken to ensure conservation of the total two-electron density within an accuracy  $10^{-8}$ during the whole duration of laser pulse.

\section{Results and discussion}
\label{sec:results}

This section consists of three subsections. In the first subsection, Sec.~\ref{sec:results_accur}, we discuss the quality of the present electronic structure calculations.  Autoionization profiles of the $2s2p\,^1P$ doubly-excited state of He, computed for different pulse durations and peak intensities, are reported and compared in Sec.~\ref{sec:results_profiles}. The obtained results are further analyzed in Sec.~\ref{sec:results_spectra} with the help of  photoelectron spectra.

\subsection{Electronic properties of system at hand}
\label{sec:results_accur}

\begin{figure}
\includegraphics[scale=1.05]{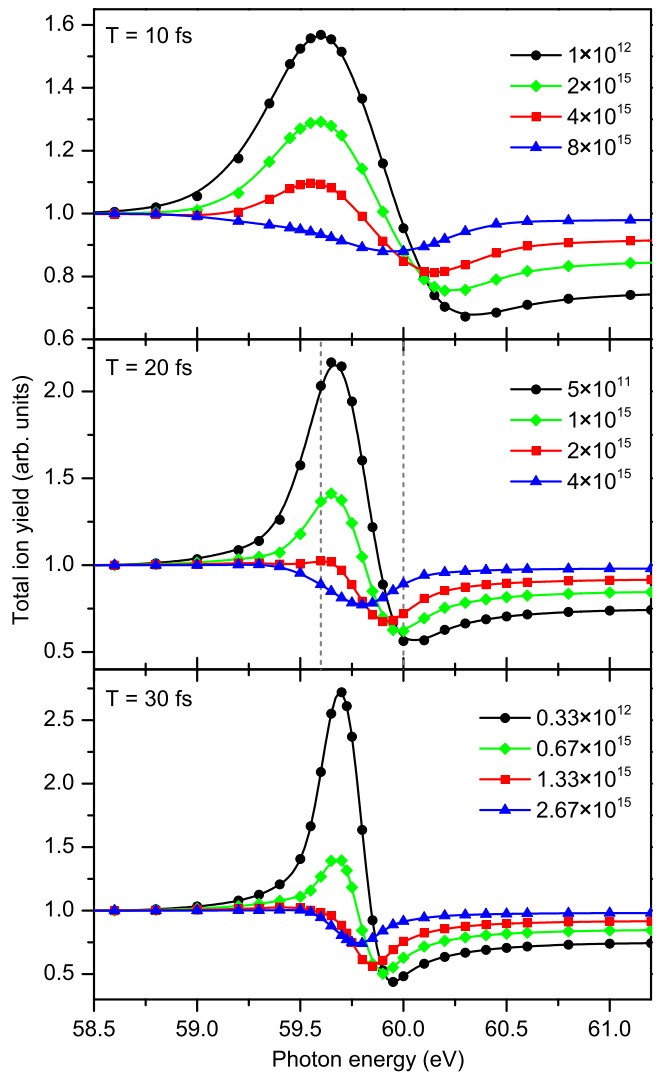}
\caption{(Color online) The total ionization yield of helium as a function of the photon energy $\omega$, computed across the $2s2p\,^1P$ doubly-excited state for different pulse durations $T$ (indicated for each panel) and peak intensities $I_0$ (indicated in the legends in W/cm$^2$). To enable comparison, all curves are shown on the relative vertical scale by normalization to unity at the photon energy of $\omega=58.5$~eV. The vertical dashed lines in the middle panel indicate the photon energies at which the spectra depicted in Fig.~\ref{fig:spectra} have been computed.}\label{fig:profile}
\end{figure}

The presently computed energy of the double-excited state of He $E(2s2p\,^1P)=-0.69319$~a.u. differs from the most accurate value of $E(2s2p\,^1P)=-0.69313$~a.u. \cite{Scrinzi98} by about 1~meV. The agreement between the presently computed $E(1s^2\,^1S)=-2.88767$~a.u. and the most accurate $E(1s^2\,^1S)=-2.90372$~a.u.~\cite{Ho91,Scrinzi98} energies of the initial ground state of He is somewhat worse (they differ by about 0.44~eV). Therefore, the presently computed resonant excitation energy    $59.71$~eV is by about 0.44~eV lower than the most accurate value of $60.15$~eV \cite{Ho91,Scrinzi98}. This difference in the photon energy should be taken into account when comparing the presently computed total ionization yields (Fig.~\ref{fig:profile}) with experimental results. Since our active space allows for an exact numerical description of the final He$^+(n\ell)$ ionic states, the presently computed ionization potential $IP=E(1s^1\,^2S)-E(1s^2\,^1S)=24.15$~eV is by about 0.44~eV lower than its experimental value of 24.59~eV \cite{NIST}. For this reason, photoelectron energies evaluated in the present calculations as $\varepsilon=\omega-IP$ are accurate and need not to be corrected when comparing the computed photoelectron spectra (Fig.~\ref{fig:spectra}) with experiment.

In order to extract autoionization lifetime of the $2s2p\,^1P$ doubly-excited state of He, we introduce the time-dependent population of all one-electron continuum states $\{\psi_\beta \}$ as:
\begin{equation}
\label{eq:Popcont}
P_c(t)=\int W(\vec{k},t)\, d^3\vec{k} ,
\end{equation}
with the three-dimensional photoelectron momentum distribution given by the Fourier transformation of the time-dependent electron wave packets
\begin{equation}
\label{eq:fourier}
W(\vec{k},t) =\frac{1}{(2\pi)^{3/2}}\sum_\beta \left\vert  \int \psi_{\beta}(\vec{r},t) \,e^{-i\vec{k}\cdot\vec{r}} d^3\vec{r}\,  \right\vert ^2.
\end{equation}
Obviously, as long as the autoionization decay of the resonant state is not essentially completed, $P_c(t)$ is smaller than the total ionization yield $P_I(T)$ (\ref{eq:probability}). Note also that $P_I(t)$ does not change in time for $t>T$. Therefore, the total photoionization probability can be considered as the upper limit of the one-electron continuum population:
\begin{equation}
\label{eq:limit}
P_I(T)=\lim_{t\to\infty} P_c(t).
\end{equation}
Finally, we introduce the following difference
\begin{equation}
\label{eq:respot}
P_r(t)=P_I(T)- P_c(t),
\end{equation}
which can be considered as the residual population of the doubly-excited state.

Numerically, we perform  dynamical calculations with a somewhat  shorter $T=5$~fs laser pulse of the moderate pulse intensity of $I_0=10^{12}$ W/cm$^2$  and the resonant carrier frequency of $\omega=59.7$~eV and evaluate the total photoionization probability $P_I(T)$ (\ref{eq:probability}) at the end of the pulse. As the next step, we additionally propagate the two-electron wave function $\Psi(\vec{r}_1,\vec{r}_2,T)$ in the absence of the laser pulse until $t=T+10$~fs and  evaluate the time-dependent population of all one-electron continuum states $P_c(t)$ (\ref{eq:Popcont}). Finally, by fitting the residual population of the doubly-excited state (\ref{eq:respot}) with the exponential-decay-law
\begin{equation}
\label{eq:decay}
P_r(t)=P_r(0)e^{-t/\tau},
\end{equation}
we extract autoionization lifetime $\tau$ at different times $t>T$. The determined lifetime converges rapidly to its final value of $\tau=16.62$~fs already at $t=10$~fs (i.e., at 5~fs after the end of the exciting pulse). This value agrees very well with the most accurate autoionization lifetimes of  $\tau=17.56$~fs \cite{Ho91} and $\tau=17.71$~fs \cite{Scrinzi98}, obtained for the $2s2p\,^1P$ doubly-excited state of He with the basis set of explicitly correlated functions. The about 6\% difference between the presently computed and the most accurate autoionization lifetimes can be explained by the use of different radial basis sets. This difference can be considered as the accuracy of the present calculations.

\subsection{Autoionization profiles}
\label{sec:results_profiles}

Being confident that the present basis provides a suitable description of the two-electron problem at hand, we proceed with the dynamics of autoionization of the $2s2p\,^1P$ state of He. Results of calculations performed for three considered pulse durations $T=10$, 20, and 30~fs  are collected respectively in the top, middle, and bottom panels of Fig.~\ref{fig:profile}. This figure depicts the total ionization yield (\ref{eq:probability}) as a function of the carrier frequency of the pulse across the $2s2p\,^1P$ resonance. For a better comparison of the autoionization profiles computed for different peak intensities of each pulse, the total ionization yields are shown in Fig.~\ref{fig:profile} on the relative scale (see figure caption for details).

Autoionization profiles obtained in the weak-field limit are shown in Fig.~\ref{fig:profile} by black curves with circles. One can see that the presently computed weak-field total ionization yields exhibit a prominent resonant feature around the resonant excitation energy of  $59.71$~eV. Apart for the discussed above 0.44~eV difference in the appearance energy of the resonance, the shape of the presently computed Fano-interference is in a good agreement with the available experimental and theoretical results (see, e.g., Refs.~\cite{MC1,Silverman64,Samson94,LaForge16,Hochstuhl12}). This shape corresponds to an asymmetric dispersion profile with large negative value of the $q$-parameter, which is additionally broadened by the spectral function $f(\omega)=\int f(t) e^{i\omega t} dt$ of the pulse.

We now turn to the strong-field limit. For the $T=20$~fs pulse, the present calculations were performed at the peak intensities of $I_0=1\times 10^{15}$, $2\times 10^{15}$ and $4\times 10^{15}$~W/cm$^2$. For the shorter $T=10$~fs and longer $T=30$~fs pulses, the three selected peak intensities were scaled to keep the product of $T\cdot I_0$ constant. One can see from Fig.~\ref{fig:profile} that, once reaching the strong-field excitation regime, increasing field intensity causes dramatic changes of the resonant profile in the computed total ionization yield. The weak-field asymmetric dispersion profiles (black curves with circles in each panel of the figure) convert their form systematically to almost symmetric dispersion profiles (green curves with rhumbuses), further on to asymmetric windows (red curves with squares), and finally to the almost symmetric transparency windows (blue curve with triangles).

\begin{figure}
\includegraphics[scale=0.9]{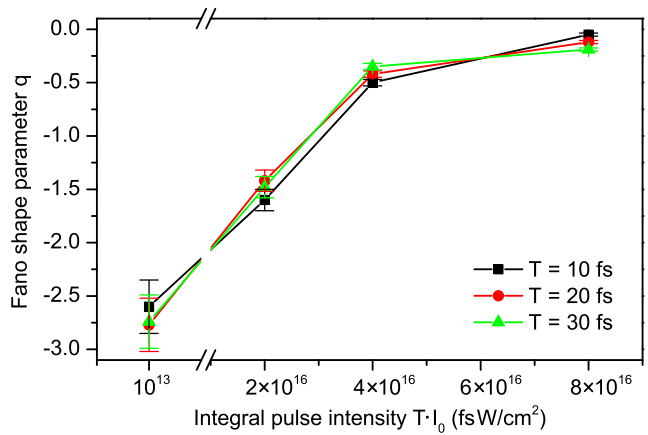}
\caption{(Color online) Fano parameter $q$ as a function of the product of pulse duration and peak intensity, $T\cdot I_0$, for pulses with different duration $T$ (see legend). The values of $q$  were extracted from the profiles depicted in Fig.~\ref{fig:profile} using parametrization \cite{Fano65,LaForge16}. The standard deviations of the fitted $q$-parameters are indicated by the vertical error bars. Note also the break in scale on the horizontal axis.}\label{fig:parameter}
\end{figure}

Such a transformation of the Fano-profile corresponds to a systematic decrease of the absolute value of the respective shape parameter $q$. In order to illustrate this fact we performed an analysis of the computed resonant profiles by applying the Fano-Cooper parametrization \cite{Fano65} (see also Eq.~(1) in Ref.~\cite{LaForge16}). The values of $q$, extracted for each pulse duration $T$, are compared in Fig.~\ref{fig:parameter} as functions of the product between pulse duration and peak intensity. One should note that, strictly seen, the Fano-Cooper parametrization \cite{Fano65,LaForge16} does not describe the situation of a short exciting laser pulse considered here, but rather applies to the case of a monochromatic continuum wave. The presently computed profiles are thus additionally broadened by the spectral function of the pulse. In addition, the Fano-Cooper ansatz assumes the interaction of an autoionizing resonance with 'flat' continua, i.e., that the direct ionization amplitudes do not vary across the resonance. In reality, the direct ionization channel in He introduces an additional asymmetry in the profile.  As a consequence, the present fitting procedure yielded rather large standard deviations of the determined $q$-parameters, which are indicated in Fig.~\ref{fig:parameter} by vertical error bars. Nevertheless, Fig.~\ref{fig:parameter} demonstrates clearly that, with the increase of the pulse intensity, the extracted parameter $q$ systematically approaches the value of zero from below.

\subsection{Photoelectron spectra}
\label{sec:results_spectra}

\begin{figure}
\includegraphics[scale=0.9]{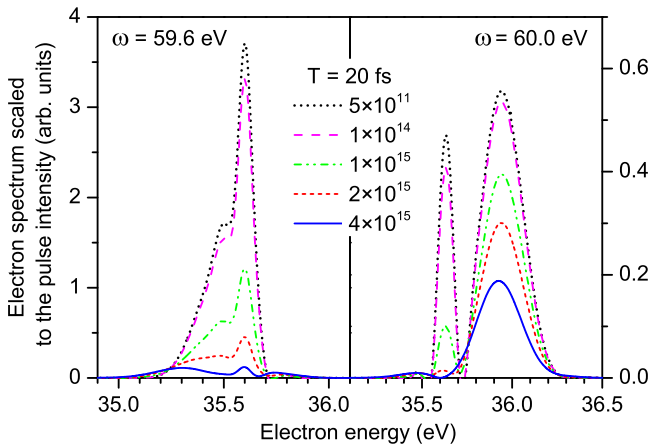}
\caption{(Color online) The photoelectron spectra of helium computed for  $T=20$~fs  laser pulses at different peak intensities (indicated in the legend in W/cm$^2$) at the photon energies of $\omega=59.6$~eV (left panel) and $\omega=60.0$~eV (right panel). Each spectrum is normalized to the corresponding peak intensity of the pulse.}\label{fig:spectra}
\end{figure}

In order to understand the presently observed trend, we made a set of calculations of the photoelectron spectrum (\ref{eq:spectrum}). These calculations were performed for two nearly-resonant exciting-photon energies of 59.6 and 60.0~eV of the $T=20$~fs laser pulse (indicated by the vertical dashed lines in the middle panel of Fig.~\ref{fig:profile}). The  photoelectron spectra computed at these two photon energies for different peak intensities of the pulse are confronted in the left and right panels of Fig.~\ref{fig:spectra}. These spectra are normalized to the respective pulse intensities (indicated in the legend). The weak-field electron spectra, obtained for the peak intensity of $5\times 10^{11}$~W/cm$^2$, are shown by dotted curves. For the peak intensities well-below $1\times 10^{14}$~W/cm$^2$ (not shown in Fig.~\ref{fig:spectra}), the intensity-normalized   spectra do not change, i.e., as expected they scale linearly with the intensity of the pulse. The present results on the linear (perturbative) regime of population of the $2s2p\,^1P$ resonance in He atom by short laser pulses are in agreement with the results reported in Ref.~\cite{Mercouris07}.

As can be recognized from Fig.~\ref{fig:spectra}, at peak intensities larger than $1\times 10^{14}$~W/cm$^2$, the relative photoelectron spectra start to deviate noticeably from the weak-field results. As the peak intensity grows, the difference between the intensity-normalized  spectra computed for each particular photon energy becomes dramatic (compare separately in each panel the dotted and solid curves). This is a clear fingerprint of the nonlinear excitation regime. Most important, these nonlinear changes are very different for the two selected photon energies. In particular, the difference between the two relative spectra computed for the smallest and the largest considered peak intensities at the photon energy of 59.6~eV (left panel of Fig.~\ref{fig:spectra}) is significantly larger than that for 60.0~eV (right panel of Fig.~\ref{fig:spectra}). This non-proportionality in change of the photoelectron spectrum is the reason for transformation of the Fano-profile in the total ionization yield, observed for the peak intensities larger than $1\times 10^{15}$~W/cm$^2$.

We mention for completeness that for peak intensities larger than $1\times 10^{15}$~W/cm$^2$, electrons released by the two-photon above-threshold ionization  process can clearly be observed in the electron spectrum around kinetic energies of 95.5~eV (not shown in Fig.~\ref{fig:spectra}). However, for the largest considered pulse intensities, contributions of the two-photon above-threshold ionization  peaks to the total ionization yields depicted in Fig.~\ref{fig:profile} do not exceed 10\% of the main photoelectron peaks shown in Fig.~\ref{fig:spectra}. Therefore, the two-photon ionization processes do not alter the non-proportional changes  demonstrated in Fig.~\ref{fig:spectra} with the help of the photoelectron spectra.

Qualitatively, the presently uncovered effects can be explained as follows. As discussed in the introduction, shape of the resonance is determined by the interplay between the amplitudes for the direct and resonant photoionization pathways (Fano-interference). In the weak-field limit, the respective probabilities scale linearly with the field strength, and the interplay between these two contributions remain unchanged. In the strong-field limit, these two contributions scale differently with the peak intensity of the pulse. As soon as the resonant photoionization pathway enter Rabi-flopping regime, its contribution stops growing. Contrary to that, contribution from the direct ionization pathway continues to increase and saturates at higher field intensities. As a consequence, the ratio of the two contributions and, thus, the absolute value of the respective $q$-parameter decrease. Similar saturation trends for the direct and resonant ionization channels were found in the resonant Auger decay of the core-excited atoms and molecules \cite{Demekhin11a,Demekhin11b,Demekhin13b,Muller15} driven by short intense coherent x-ray pulses.

\section{Conclusion}
\label{sec:summary}

The interaction of a helium atom with an intense short high-frequency laser pulse is studied theoretically by the time-dependent restricted-active-space configuration-interaction method in the dipole velocity gauge. The present active space of two-electron configurations permits only one of the electrons in He to be ionized. This photoelectron was represented on the radial grid by the time-dependent wave packets with angular momenta $\ell\leq 4$. The active bound electron was allowed to populate discrete states $\left\{n\ell_+\right\}$ of the He$^+$ ion with $n\leq 5$ and $\ell\leq 4$. In the calculations, the carrier frequency of the pulse was scanned across the excitation energy of the $2s2p\,^1P$ doubly-excited state of He. The pulse durations were chosen to be comparable with the autoionization lifetime of the resonance, whereas the corresponding pulse intensities were made sufficiently large to initiate Rabi-flopping between the initial ground and the intermediate resonant electronic states.

We demonstrate that an intense exciting laser pulse can substantially modify the Fano-interference between the amplitudes for the direct photoionization and for the excitation and decay of the intermediate resonance, both populating the same final continuum state. In the nonlinear excitation regime, relative contributions from the two photoionization pathways scale differently with the peak intensity of the pulse: The direct ionization channel saturates at much larger pulse strengths than the resonant one. As a consequence, the absolute value of the corresponding Fano-parameter $q$ decreases as the pulse intensity grows. At sufficiently large pulse strengths, the asymmetric dispersion profile of the total ionization yield in the weak-field regime transforms into an almost symmetric transparency-window Fano-profile.

Although we are interested here in pulses of similar duration as the autoionization lifetime, we would like to mention that interesting effects can also be expected when employing pulses with completely different durations. For instance, if the pulse is very short, the $q$ parameter would not necessarily converge to zero with intensity as it is the case in the present study.

Our work provides a clear numerical demonstration of how the Fano-interference can be controlled by a single intense exciting laser pulse. This scheme does not require a second probe pulse, like, e.g., in the pump-probe experiments performed in Ref.~\cite{Ott13}, where control over the Fano-interference was achieved through the phase of a time-dependent dipole-response function. Finally, we notice that the carrier frequencies, pulse durations, field intensities, and temporal coherence, required to produce observable effects in He, are available at the free electron laser facility FERMI@Elettra \cite{FERMI}. At present, it generates single coherent pulses in the photon energy range of 12 to 413~eV with  durations of 30 to 100 fs. It is expected to provide a flux of up to $10^{14}$ photons per pulse, which combined with the appropriate focusing optics \cite{Sorokin07}, enables to access irradiance of about $10^{16}$~W/cm$^2$. Our numerical results provide a theoretical background for experimental verification of this effect at presently available high-frequency laser pulse facilities.

\begin{acknowledgements}
This work was partly supported by the F\"{o}rderprogramm zur weiteren Profilbildung in der Universit\"{a}t Kassel (F\"{o}rderlinie \emph{Gro{\ss}e Br\"{u}cke}), by the DFG project No. DE 2366/1-1, and by the U.S. ARL and the U.S. ARO  Grant No. W911NF-14-1-0383. The authors would like to thank Thomas Pfeifer for many valuable discussions.
\end{acknowledgements}


\begin{thebibliography}{99}

\bibitem{Beutler35}
H. Beutler, Z. Phys. \textbf{93}, 177 (1935).

\bibitem{Fano35}
U. Fano, Nuovo Cimento \textbf{12}, 154 (1935).

\bibitem{MC1}
R.P. Madden and K. Codling, Phys. Rev. Lett. \textbf{10}, 516 (1963).

\bibitem{MC2}
R.P. Madden and K. Codling, J. Opt. Soc. Am. \textbf{54}, 268 (1964).

\bibitem{MC3}
R.P. Madden and K. Codling,  Astrophys. J. \textbf{141}, 364 (1965).

\bibitem{Sukhorukov12}
V.L. Sukhorukov, I.D. Petrov, M. Sch\"{a}fer, F. Merkt, M.-W. Ruf, and H. Hotop, J. Phys. B \textbf{45}, 092001 (2012).

\bibitem{Fano61}
U. Fano, Phys. Rev. \textbf{124}, 1866 (1961).

\bibitem{Fano65}
U. Fano and J.W. Cooper, Phys. Rev. \textbf{137}, A1364 (1965).

\bibitem{Shore67}
B.W. Shore, J. Opt. Soc. Am. \textbf{57}, 881 (1967).

\bibitem{Shore68}
B.W. Shore,  Phys. Rev. \textbf{171}, 43 (1968).

\bibitem{Lambropoulos80}
P. Lambropoulos, Appl. Opt. \textbf{19}, 3926 (1980).

\bibitem{Lambropoulos98}
P. Lambropoulos, P. Maragakis, and J. Zhang, Physics Reports \textbf{305}, 203 (1998).

\bibitem{Rzaznewski81}
K. Rz\c{a}\.{z}newski and J.H. Eberly, Phys. Rev. Lett. \textbf{47}, 408 (1981).

\bibitem{Rzaznewski83}
K. Rz\c{a}\.{z}newski, Phys. Rev. A \textbf{28}, 2565 (1983).

\bibitem{Toyota07}
K. Toyota, O.I. Tolstikhin, T. Morishita, and S. Watanabe, Phys. Rev. A \textbf{76}, 043418 (2007).

\bibitem{Toyota08}
K. Toyota, O.I. Tolstikhin, T. Morishita, and S. Watanabe, Phys. Rev. A \textbf{78}, 033432 (2008).

\bibitem{Demekhin12a}
Ph.V. Demekhin and L.S. Cederbaum, Phys. Rev. Lett. \textbf{108}, 253001 (2012).

\bibitem{Demekhin12b}
Ph.V. Demekhin and L.S. Cederbaum, Phys. Rev. A \textbf{86}, 063412 (2012).

\bibitem{Baghery17}
M. Baghery, U. Saalmann, and J.-M. Rost, Phys. Rev. Lett.  \textbf{118}, 143202 (2017).

\bibitem{Lambropoulos81}
P. Lambropoulos and P. Zoller, Phys. Rev. A \textbf{24}, 379 (1981).

\bibitem{Krausz09}
F. Krausz and M. Ivanov, Rev. Mod. Phys. \textbf{81}, 163 (2009).

\bibitem{Sansone06}
G. Sansone, E. Benedetti, F. Calegari \emph{et al.}, Science \textbf{314}, 443 (2006).

\bibitem{Goulielmakis08}
E. Goulielmakis, M. Schultze, M. Hofstetter \emph{et al.}, Science \textbf{320}, 1614 (2008).

\bibitem{Ackermann07}
W. Ackermann, G. Asova, V. Ayvazyan \emph{et al.}, Nature Photon. \textbf{1}, 336 (2007).

\bibitem{FERMI}
Home page of FERMI at Elettra in Trieste, Italy, www.elettra.trieste.it/lightsources/fermi/machine.html

\bibitem{FLASH}
J.T. Costello, J. Phys.: Conf. Series \textbf{88}, 012057 (2007).

\bibitem{Fushitani08}
M. Fushitani, Annu. Rep. Prog. Chem. Sect. C: Phys. Chem. \textbf{104}, 272 (2008).

\bibitem{Wang10}
H. Wang, M. Chini, S. Chen, C.-H. Zhang, F. He, Y. Cheng, Y. Wu, U. Thumm, and Z. Chang, Phys. Rev. Lett. \textbf{105}, 143002 (2010).

\bibitem{Loh08}
Z.-H. Loh, C.H. Greene, and S.R. Leone, Chem. Phys. \textbf{350}, 7 (2008).

\bibitem{Gaarde11}
M.B. Gaarde, C. Buth, J.L. Tate, and K.J. Schafer, Phys. Rev. A \textbf{83}, 013419 (2011).

\bibitem{Holler11}
M. Holler, F. Schapper, L. Gallmann, and U. Keller, Phys. Rev. Lett. \textbf{106}, 123601 (2011).

\bibitem{Chen12}
S. Chen, M.J. Bell, A.R. Beck, H. Mashiko, M. Wu, A.N. Pfeiffer, M.B. Gaarde, D.M. Neumark, S.R. Leone, and K.J. Schafer, Phys. Rev. A \textbf{86}, 063408 (2012).

\bibitem{Ott13}
C. Ott, A. Kaldun, P. Raith, K. Meyer, M. Laux, J. Evers, C.H. Keitel, C.H. Greene, T. Pfeifer, Science \textbf{340}, 716 (2013).

\bibitem{Chini14}
M. Chini, X.Wang, Y. Cheng, and Z. Chang, J. Physics B \textbf{47}, 124009 (2014).

\bibitem{Blattermann14}
A. Bl\"{a}ttermann, C. Ott, A. Kaldun, T. Ding, and T. Pfeifer, J. Physics B \textbf{47}, 124008 (2014).

\bibitem{Kaldun14}
A. Kaldun, C. Ott, A. Bl\"{a}ttermann, M. Laux, K. Meyer, T. Ding, A. Fischer, and T. Pfeifer, Phys. Rev. Lett. \textbf{112}, 103001 (2014).

\bibitem{Ott14}
C. Ott, A. Kaldun, L. Argenti, P. Raith, K. Meyer, M. Laux, Y. Zhang, A. Bl\"{a}ttermann, S. Hagstotz, T. Ding, R. Heck, J. Madroñero, F. Mart\'{\i}n	 and T. Pfeifer, Nature \textbf{516}, 374 (2014).

\bibitem{Chen13a}
S. Chen, M. Wu, M.B. Gaarde, and K.J. Schafer, Phys. Rev. A \textbf{87}, 033408 (2013).

\bibitem{Chen13b}
S. Chen, M. Wu, M.B. Gaarde, and K.J. Schafer, Phys. Rev. A \textbf{88}, 033409 (2013).

\bibitem{Argenti15}
L. Argenti, \'{A} Jim\'{e}nez-Gal\'{a}n, C. Marante, C. Ott, T. Pfeifer, and F. Mart\'{\i}n, Phys. Rev. A \textbf{91}, 061403 (2015).

\bibitem{Silverman64}
S.M. Silverman and E.N. Lassettre,  J. Chem. Phys. \textbf{40}, 1265 (1964).

\bibitem{Samson94}
J. Samson, Z. He, L. Yin, and A. Haddad, J. Phys. B \textbf{27}, 887 (1994).

\bibitem{LaForge16}
A.C. LaForge, D. Regina, G. Jabbari, K. Gokhberg, N.V. Kryzhevoi, S.R. Krishnan, M. Hess, P. O'Keeffe, A. Ciavardini, K.C. Prince, R. Richter, F. Stienkemeier, L.S. Cederbaum, T. Pfeifer, R. Moshammer, and M. Mudrich, Phys. Rev. A \textbf{93}, 050502(R) (2016).

\bibitem{Hochstuhl12}
D. Hochstuhl and M. Bonitz, Phys. Rev. A \textbf{86}, 053424 (2012).

\bibitem{Artemyev16}
A.N. Artemyev, A.D. M\"{u}ller, D. Hochstuhl, L.S. Cederbaum, and Ph.V. Demekhin, Phys. Rev. A \textbf{93}, 043418 (2016).

\bibitem{Artemyev17}
A. N. Artemyev, L. S. Cederbaum, and P. V. Demekhin, hys. Rev. A  \textbf{95}, 033402 (2017).

\bibitem{Cormier96}
E. Cormier and P. Lambropoulos, J. Phys. B \textbf{29}, 1667 (1996).

\bibitem{Han10}
Y.-C. Han and L.B. Madsen, Phys. Rev. A \textbf{81}, 063430 (2010).

\bibitem{Hochstuhl11}
D. Hochstuhl and M. Bonitz, J. Chem. Phys. \textbf{134}, 084106 (2011).

\bibitem{Manolopoulos88}
D.E. Manolopoulos and R.E. Wyatt, Chem. Phys. Lett. \textbf{152}, 23 (1988).

\bibitem{Rescigno00}
T.N. Rescigno and C.W. McCurdy, Phys. Rev. A \textbf{62}, 032706 (2000).

\bibitem{McCurdy04}
C.W. McCurdy, M. Baertschy and T.N. Rescigno, J. Phys. B \textbf{37}, R137 (2004).

\bibitem{Demekhin13a}
Ph.V. Demekhin, D. Hochstuhl,  and  L.S. Cederbaum, Phys. Rev. A \textbf{88}, 023422 (2013).

\bibitem{Artemyev15}
A.N. Artemyev, A.D. M\"{u}ller, D. Hochstuhl, and Ph.V. Demekhin, J. Chem. Phys. \textbf{142}, 244105 (2015).

\bibitem{Park86}
T.J. Park and J.C. Light, J. Chem. Phys. \textbf{85}, 5870 (1986).

\bibitem{Ho91}
Y.K. Ho, Z. Phys. D  \textbf{21}, 191 (1991).

\bibitem{Scrinzi98}
A. Scrinzi and B. Piraux, Phys. Rev. A \textbf{58}, 1310 (1998).

\bibitem{NgokoDjiokap11}
J.M. Ngoko Djiokap and A.F. Starace, Phys. Rev. A \textbf{84}, 013404 (2011).

\bibitem{Polizzi09}
E. Polizzi, Phys. Rev. B \textbf{79}, 115112 (2009).

\bibitem{Polizzi15}
E. Polizzi and J. Kestyn (2015) arXiv:1203.4031v3.

\bibitem{Bandrauk09}
A.D. Bandrauk, S. Chelkowski, D.J. Diestler, J. Manz, and K.-J. Yuan, Phys. Rev. A \textbf{79}, 023403 (2009).

\bibitem{Krause92}
J.L. Krause, K.J. Schafer, and K.C. Kulander, Phys. Rev. A \textbf{45}, 4998 (1992).

\bibitem{NIST}
A. Kramida, Y. Ralchenko, and J. Reader, NIST Atomic Spectra Database (National Institute of Standards and Technology, Gaithersburg, MD, 2012), http://physics.nist.gov/PhysRefData/ASD/index.html.

\bibitem{Mercouris07}
Th. Mercouris, Y. Komninos, and C.A. Nicolaides, Phys. Rev. A \textbf{75}, 013407 (2007).

\bibitem{Demekhin11a}
Ph.V. Demekhin and L.S. Cederbaum, Phys. Rev. A \textbf{83}, 023422 (2011).

\bibitem{Demekhin11b}
Ph.V. Demekhin, Y.-C. Chiang, and L.S. Cederbaum, Phys. Rev. A \textbf{84}, 033417 (2011).

\bibitem{Demekhin13b}
Ph.V. Demekhin and L.S. Cederbaum, J. Phys. B \textbf{46}, 164008 (2013).

\bibitem{Muller15}
A.D. M\"{u}ller and Ph.V. Demekhin, J. Phys. B \textbf{48}, 075602 (2015).

\bibitem{Sorokin07}
A.A. Sorokin, S.V. Bobashev, T. Feigl, K. Tiedtke, H. Wabnitz, and M. Richter, Phys. Rev. Lett. \textbf{99}, 213002 (2007).




\end{thebibliography}
\end{document}